\documentclass[final,1p,times]{elsarticle}
\usepackage{graphicx}
\usepackage{amssymb}
\usepackage{amsthm}

\journal{Nuclear Physics A} 

\begin{document}


\begin{frontmatter} 


\title{Bulk viscosity effects on elliptic flow}

\author{G. S. Denicol${}^{a}$, T. Kodama${}^{b}$, T. Koide${}^{c}$, and Ph.
Mota${}^{b}$}

\address{$^{a}$Institute f\"ur Theoretische Physik, 
Johann Wolfgang Goethe-Universit\"at, Max-von-Laue Str. 1,
60438, Frankfurt am Main, Germany }
\address{$^{b}$Instituto de F\'{\i}sica, Universidade Federal do Rio de Janeiro, C.
P. 68528, 21945-970, Rio de Janeiro, Brasil}
\address{$^{c}$FIAS, Johann Wolfgang Goethe-Universit\"at, Ruth-Moufang Str. 1,
60438, Frankfurt am Main, Germany}

\begin{abstract} 
The effects of bulk viscosity on the elliptic flow $v_{2}$ are studied using
realistic equation of state and realistic transport coefficients. We find that the
bulk viscosity acts in a non trivial manner on $v_{2}$.  At low $p_{T}$ , the reduction 
of $v_{2}$ is even more effective compared to the case of shear
viscosity,  whereas at high $p_{T}$,  an enhancement  of $v_{2}$ compared to the ideal
case is observed.  We argue that this is caused by the competition of the
critical behavior of the equation of state and the transport coefficients.
\end{abstract} 

\end{frontmatter}



\section{Introduction}

The effect of shear viscosity in heavy ion collisions has been studied by
several authors \cite{Romatschke1} and was found to affect
considerably the elliptic flow observable $v_{2}$. On the other hand, the
bulk viscosity has always been neglected and its effects are mostly unknown.
The shear viscosity acts as the resistance against the deformation of a
fluid element. In this sense, it is natural to expect that $v_{2}$, which
characterizes the spatial anisotropy of the dynamics, is affected by the
shear viscosity. On the other hand, the bulk viscosity acts only against the
expansion or compression of the fluid. However, this does not mean that the
bulk viscosity has negligible effects on $v_{2}$, particularly because the
expansion and compression of the produced matter in the collisions is also
anisotropic.

Therefore, the behavior of the bulk viscosity coefficient, $\zeta $%
, near the QCD phase transition should also be carefully studied. Recently,
lattice QCD simulations \cite{Karsch} suggest that $\zeta $ is strongly
enhanced near the critical temperature $T_{c}$. A similar enhancement will
appear even below $T_{c}$, as is shown in \cite{Noronha}, where $\zeta $ in
the hadronic phase increases also quickly towards $T_{c}$. Considering the
fact that the fluid dynamics slows down near $T_{c}$ due to the small sound
velocity $c_{s}$, the critical behavior of $\zeta $ should have a crucial
role on the expansion of the matter.

In this work , we investigate the effect of bulk viscosity on $v_{2}$ by
incorporating the critical behavior of the equation of state (EoS) and $%
\zeta $ within a 2+1 dimensional hydrodynamic modeling. On the other hand,
the dissipative corrections to the one-particle distribution function at
freeze out is not considered, because Grad's method is not applicable for
the freeze out values of bulk viscosity obtained in our calculations \cite%
{paper}.

\section{Relativistic Dissipative Hydrodynamics}

\label{sec2}

The formulation of relativistic dissipative hydrodynamics is not trivial.
The naive relativistic generalization of the Navier-Stokes equation violates
causality and leads to intrinsic instabilities making the theory not
applicable. In this work, we use the memory function method \cite{dkkm,dkkm4}%
, which is formulated in such a way that the magnitude of the bulk viscosity
has a lower bound, and it is possible to implement stable numerical
simulations even for ultra-relativistic initial conditions.

We consider the case of vanishing baryon chemical potential where only the
conservation of the energy and momentum is required. For a general metric $%
g_{\mu \nu }$, the continuity equation for the energy-momentum
tensor is given by 
\begin{equation}
\frac{1}{\sqrt{-g}}\partial _{\mu }\left( \sqrt{-g}T^{\mu \nu }\right)
+\Gamma _{\lambda \mu }^{\nu }T^{\lambda \mu }=0,  \label{conserv}
\end{equation}%
where $\Gamma _{\mu \lambda }^{\nu }$ is the Christoffel symbol and $g$ is
the determinant of $g_{\mu \nu }$. We use the Landau definition for the
local rest frame and assume, as usual, that the thermodynamic relations are
valid in this frame. We further ignore the shear viscosity. Then the
energy-momentum tensor is expressed as $T^{\mu \nu }=\left( \varepsilon
+p+\Pi \right) u^{\mu }u^{\nu }-\left( p+\Pi \right) g^{\mu \nu }$, where $%
\varepsilon $, $p$, $u^{\mu }$ and $\Pi $ are the energy density, pressure,
four velocity and bulk viscosity, respectively. The equation for the entropy
production is, 
\begin{equation}
\frac{1}{\sqrt{-g}}\partial _{\mu }\left( \sqrt{-g}su^{\mu }\right) =-\frac{%
\Pi }{T}\frac{1}{\sqrt{-g}}\partial _{\mu }\left( \sqrt{-g}u^{\mu }\right) .
\label{entrop_conserv}
\end{equation}%
where $s$ is the entropy density. From this, we can define the thermodynamic
force $F$ associated with the bulk viscosity as $\sqrt{-g}^{-1}\partial
_{\mu }\left( \sqrt{-g}u^{\mu }\right) $. It is also useful to introduce
the specific volume $\sigma $ defined by the conservation law $\sqrt{-g}%
^{-1}\partial _{\mu }\left( \sqrt{-g}\sigma u^{\mu }\right) =0$.

In the memory function method, an irreversible current $J$ is induced by the
corresponding thermodynamic force as 
\begin{equation}
\frac{J(\tau )}{\sigma (\tau )}=-\int_{\tau _{0}}^{\tau }d\tau ^{\prime }%
\frac{1}{\tau _{R}(\tau ^{\prime })}\exp {\left( -\int_{\tau ^{\prime
}}^{\tau }\frac{d\tau ^{\prime \prime }}{\tau _{R}(\tau ^{\prime \prime })}%
\right) }\frac{F(\tau ^{\prime })}{\sigma (\tau ^{\prime })}+\frac{J(\tau
_{0})}{\sigma (\tau _{0})}\exp {\left( -\int_{\tau _{0}}^{\tau }\frac{d\tau
^{\prime }}{\tau _{R}(\tau ^{\prime })}\right) }.
\end{equation}%
Thus, the equation of the bulk viscosity is obtained by setting $J=\Pi $ and 
$F=\sqrt{-g}^{-1}\partial _{\mu }\left( \sqrt{-g}u^{\mu }\right) $. This is
re-expressed in a differential form as follows, 
\begin{equation}
\tau _{\mathrm{R}}u^{\mu }\partial _{\mu }\left( \frac{\Pi }{\sigma }\right)
+\frac{\Pi }{\sigma }=-\frac{\zeta }{\sigma }\frac{1}{\sqrt{-g}}\partial
_{\mu }\left( \sqrt{-g}u^{\mu }\right) ,  \label{GeneralBulk}
\end{equation}%
where $\tau _{\mathrm{R}}$ is the relaxation time and $\zeta $ is the bulk
viscosity coefficient.

In the following calculations, we consider the hyperbolic coordinate system $%
x^{\mu }=(\tau ,\mathbf{r}_{T},\eta )$, where $\tau =\sqrt{t^{2}-z^{2}}$,$%
~~~\eta =\frac{1}{2}\tanh \left( \frac{t+z}{t-z}\right) $, $~~\mathbf{r}%
_{T}=(x,y)$ and $\sqrt{-g}=\tau $.

\begin{figure}[tbp]
\begin{minipage}{.5\linewidth}
\includegraphics[scale=0.25]{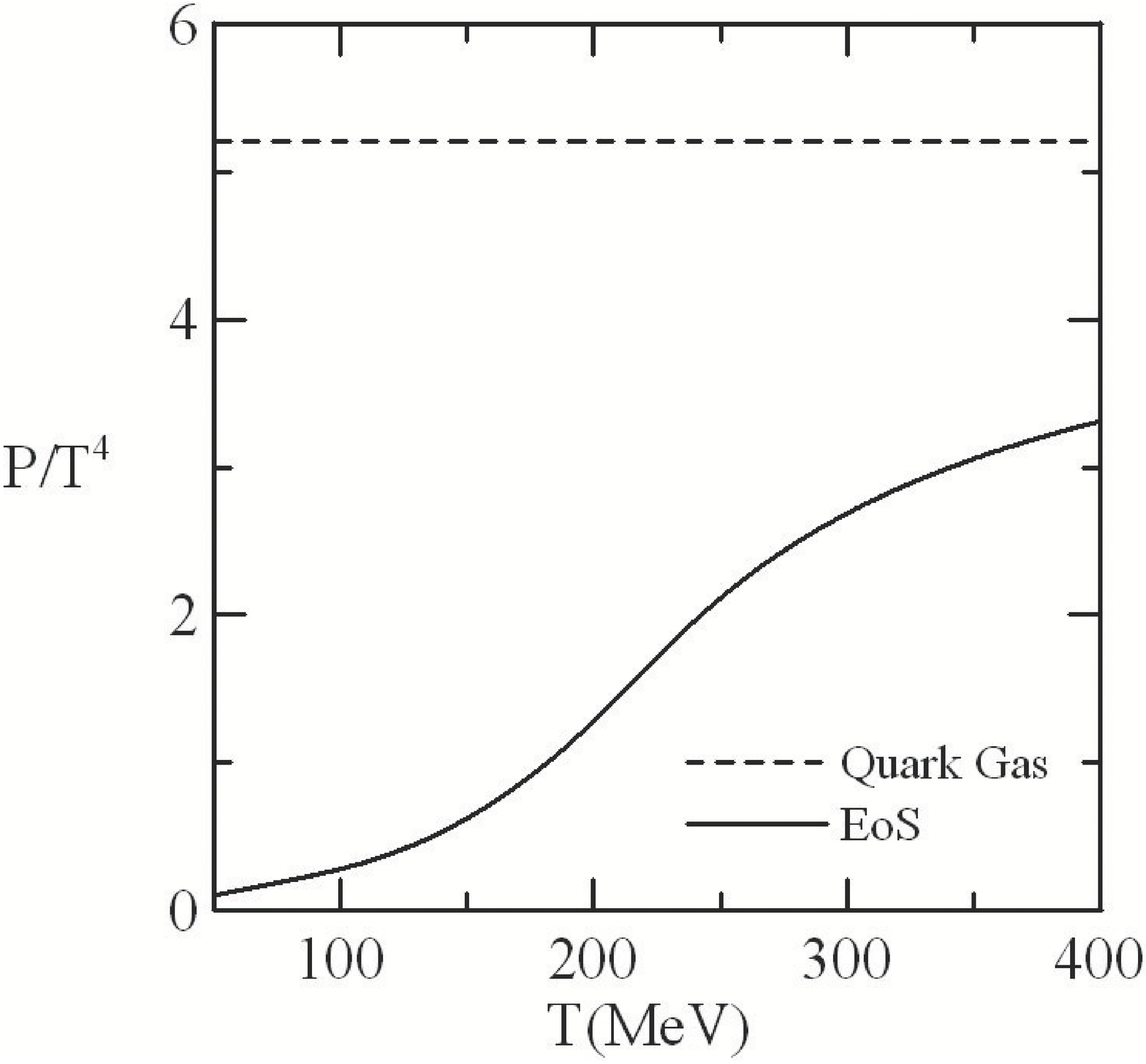}
\end{minipage}
\begin{minipage}{.5\linewidth}
\includegraphics[scale=0.25]{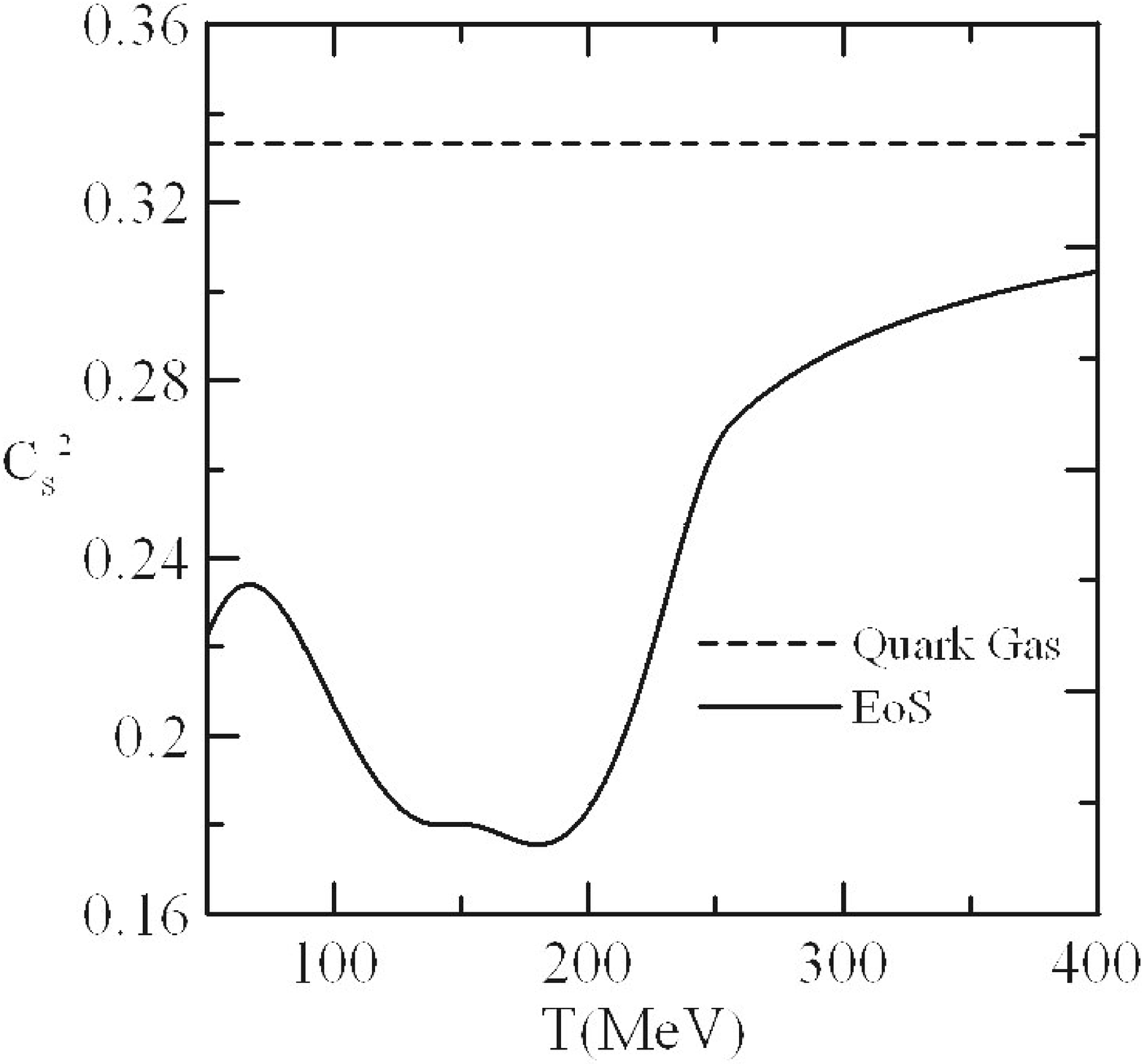}
\end{minipage}
\caption{The temperature dependence of the EoS. The solid line
corresponds to the EoS used in our calculations. The dotted line
denotes the ideal gas of massless quarks EoS.}
\label{eos}
\end{figure}

To solve numerically the relativistic hydrodynamic equations (\ref{conserv}%
), (\ref{entrop_conserv}) and (\ref{GeneralBulk}), we use the Smoothed
Particle Hydrodynamic method \cite{paper, dkkm4, dkkm2}. The initial
condition used is a parametrization of the usual Glauber model. The form of
this initial condition and the choice of parameters can be found in \cite%
{paper}. The impact parameter $b$ is taken to be $7$ fm. For the EoS, we
consider a smooth interpolation between the lattice QCD results \cite%
{AokiEoS} and the hadron resonance gas, which is shown in Fig. \ref{eos}.

For the bulk viscosity coefficient, we use the result from lattice QCD \cite%
{Karsch} for the QGP phase and the result from \cite{Noronha} for the hadron
phase. Then, the ratio $\zeta /s$ exhibits a maximum at $T_{c}$ and starts
to decrease almost exponentially as the system departs from $T_{c}$.
We express these features of $\zeta /s$ as 
\begin{equation}
\frac{\zeta }{s}=\left\{ 
\begin{array}{ll}
A_{1}x^{2}+A_{2}x-A_{3} & (0.995T_{c}\geq T\geq 1.05T_{c}) \\ 
\lambda _{1}\exp (-(x-1)/\sigma _{1})+\lambda _{2}\exp (-(x-1)/\sigma
_{2})+0.001 & (T>1.05T_{C}) \\ 
\lambda _{3}\exp ((x-1)/\sigma _{3})+\lambda _{4}\exp ((x-1)/\sigma
_{4})+0.03 & (T<0.995T_{C}),%
\end{array}%
\right. 
\end{equation}%
where $x=T/T_{C}$. The fitted parameters are $\lambda _{1}=\lambda _{3}=0.9$%
, $\lambda _{2}=0.25$, $\lambda _{4}=0.22$, $\sigma _{1}=10\sigma _{3}=0.025$%
, $\sigma _{4}=0.022,$ $A_{1}=-13.77$, $A_{2}=27.55$ and $A_{3}=13.45$. For
the relaxation time, we use the following parameterization $\tau _{R}=b_{\Pi
}\zeta /(\epsilon +p)$ \cite{dkkm4}. In this work, we use $b_{\Pi }=6$ \ 
\cite{paper}.

\section{Numerical results}

\label{sec4}

\begin{figure}[tbp]
\centering \includegraphics[scale=0.25]{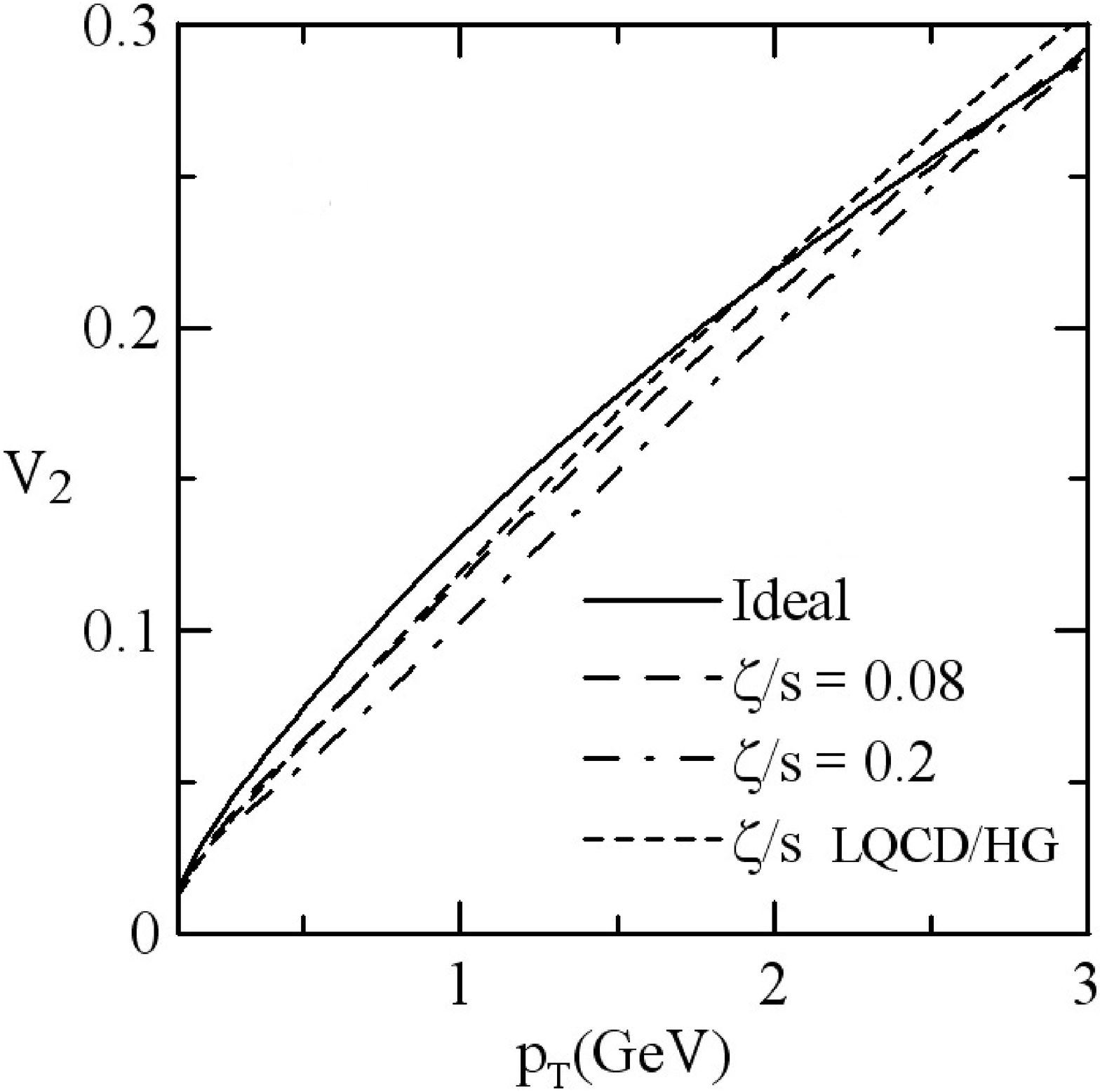}
\caption{$v_{2}$ as a function of $p_{T}$. The figure displays
results for different choices of $\protect\zeta /s$.
The solid line corresponds to the ideal result, the dashed and dash-dotted
lines correspond respectively to $\protect\zeta /s=$ $0.08$ and $0.2$, and
the dotted line is the result for $\protect\zeta /s$ with critical behavior.}
\label{Lattice}
\end{figure}

We calculate the elliptic flow parameter, $v_{2}=\left\langle \cos (2\phi
)\right\rangle $, for pions as a function of the transverse momentum, using
the Cooper-Frye method with a freeze out temperature $130$ MeV \cite{paper}.
In Fig. \ref{Lattice}, $v_{2}$ is shown for viscous fluids (dashed,
dash-dotted and dotted lines) and ideal fluid (solid line). Dashed and
dash-dotted lines correspond to the constant $\zeta $/$s$ cases and are
shown just for the sake of comparison, whereas the dotted line represents
the $\zeta /s$ with the QCD phase transition.

For small transverse momenta $P_{T}$, $v_{2}$ is suppressed by the effect of
bulk viscosity, similar to the case of shear viscosity. However, at high $%
P_{T}$, $v_{2}$ starts to recover and even surpasses the values obtained
from ideal hydrodynamics. This effect is enhanced when $\zeta /s$
exhibits a critical behavior. This behavior at high $P_{T}$ is not observed for the case
of shear viscosity and we further confirmed that this is not observed when an
ideal gas of massless quarks EoS is used \cite{paper}. Thus, this effect should be attributed to the presence of the QCD phase transition, both in the EoS and in the bulk viscosity.  

This behavior can be understood as follows. For the viscous case, the 
radial dependence of the fluid flow is not monotonic in the region
where the velocity of sound is minimum. This is because the flow of the
internal matter, which has a higher temperature, tends to catch up the
foregoing fluid elements, generating an inhomogeneous velocity field in the
radial direction. Then, the bulk
viscosity $\Pi $ starts to increase in this region, heating the
matter. If this happens, the gradient of $\Pi $ becomes dominant in the
acceleration of the fluid compared to the pressure gradient (note that the
acceleration is given by $-\nabla \left( p+\Pi \right) $ ) since the
gradient of the pressure is proportional to the sound velocity. Such a mechanism works more efficiently in the direction where the initial
acceleration is large, and in consequence, in the direction to increase the
elliptic flow. Furthermore, this effect becomes effective only when
significant collective flow is formed near T$_{c}$. Thus the recovery of $%
v_{2}$ by this mechanism is expected for higher momentum particles.

We acknowledge illuminating discussions with T. Hirano and A. Monnai. This
work has been supported by CNPq, FAPERJ, CAPES, PRONEX and the Helmholtz
International Center for FAIR within the framework of the LOEWE program
(Landesoffensive zur Entwicklung Wissenschaftlich- Okonomischer Exzellenz)
launched by the State of Hesse.




\end{document}